\documentclass[aps, twocolumn, preprint secnumarabic, nobalancelastpage, nofootinbib]{revtex4-2} 

\usepackage{graphicx}
\usepackage{dcolumn}
\usepackage[dvipsnames]{xcolor}
\usepackage{bm}
\usepackage{hyperref}
\usepackage{float}
\usepackage{upgreek}
\usepackage{amsmath}
\usepackage{amssymb}
\hypersetup{
    colorlinks=true,
    linkcolor=blue,
    filecolor=magenta,      
    urlcolor=cyan,
    pdftitle={Overleaf Example},
    pdfpagemode=FullScreen,
    }
\newcommand\scalemath[2]{\scalebox{#1}{\mbox{\ensuremath{\displaystyle #2}}}}
    
\begin{document}

\preprint{APS/123-QED}

\title{Kardar-Parisi-Zhang physics in optically-confined continuous polariton condensates}

\author{Mikhail Misko$^{1,\bullet}$, Natalia Starkova$^{1,\bullet}$, and Pavlos G. Lagoudakis$^{1}$ }
\thanks{Correspondence address: p.lagoudakis@skol.tech\\$^{\bullet}$ M.M. and N.S. contributed equally to this work}

\affiliation{$^1$Hybrid Photonics Laboratory, Skolkovo Institute of Science and Technology, Territory of Innovation Center Skolkovo, Bolshoy Boulevard 30, building 1, 121205 Moscow, Russia}

\date{\today}

\begin{abstract}
Kardar–Parisi–Zhang (KPZ) scaling has been observed in discrete polariton lattices, enabled by engineered band structures that stabilize the condensate. Whether this universality extends to intrinsically continuous systems with natural noise regularization remains an open question. We propose and numerically demonstrate KPZ scaling in a continuous quasi-one-dimensional polariton condensate stabilized by optical confinement in the transversal direction. Large-scale simulations of the stochastic Gross–Pitaevskii equation, with experimentally relevant parameters, reveal temporal and spatial scaling exponents of the two-point phase correlation function \(\beta_C \approx 0.30(5)\) and \(\alpha_C \approx 0.46(8)\), and Tracy-Widom one-point phase fluctuation statistics, yielding robust KPZ dynamics intrinsic to the continuous polariton fluid.
\end{abstract}

\maketitle

Many complex stochastic systems display universal large-scale behavior characterized by only a few scaling exponents that are independent of the microscopic interaction details and are defined only by the system's symmetries and dimensionality. A central example is the Kardar–Parisi–Zhang (KPZ) universality class of growing interfaces, whose coarse-grained description is given by the KPZ equation for a fluctuating height field \cite{kardar1986dynamic}. Despite its apparent simplicity, the KPZ equation is a singular stochastic partial differential equation, it lacks the continuous limit as the discretization step tends to zero. Its analytical solution and numerical integration are highly non-trivial, requiring renormalization techniques \cite{dupuis2021nonperturbative} and careful discretization schemes \cite{hairer2013solving}, respectively. Moreover, the asymptotic nature of KPZ physics demands for large effective size of the system, which can roughly be defined as the ratio between the system size and noise correlation length. This motivates the search for physical platforms that naturally regularize KPZ dynamics by providing the smallest possible UV noise cutoff, and allow direct access to universal exponents, correlation functions, and fluctuation statistics, thereby providing analog simulators of non-linear, non-equilibrium universality classes.

Exciton–polariton condensates in semiconductor microcavities are particularly attractive in this context. They are driven–dissipative quantum fluids of light–matter with strong interactions, controllable pumping and loss, and direct optical access to both amplitude and phase of the order parameter \cite{carusotto2013quantum}. A wide range of non-equilibrium phenomena have been realised with polaritons including condensation \cite{imamog1996nonequilibrium, kasprzak2006bose}, superfluidity \cite{amo2009superfluidity}, vortex formation \cite{lagoudakis2008quantized, sanvitto2010persistent}, turbulence \cite{koniakhin20202d, ferrini2025driven} and topological phase transitions\cite{caputo2018topological}. Recent theoretical efforts investigated polariton realizations of KPZ universality under resonant and non-resonant pumping in 1D and 2D polariton condensates \cite{he2015scaling, altman2015two, zamora2017tuning, zamora2020vortex,squizzato2018kardar, deligiannis2021accessing, deligiannis2022kpz, vercesi2023phase, helluin2025phase}.

In the pioneering work of Fontaine \emph{et al.} \cite{fontaine2022kardar}, KPZ exponents and Tracy–Widom statistics were reported in a one-dimensional Lieb lattice of quantum-well micropillars, where band-structure engineering enables condensation into an effective negative-mass state and suppresses modulational instabilities\cite{baboux2018unstable}. More recently, Widmann \emph{et al.} demonstrated KPZ scaling in a two-dimensional micropillar array with tailored dispersion \cite{widmann2025observation}. In both cases, however, the explicitly discretized lattice geometry and engineered band structure are integral to stabilizing the condensate and suppressing vortices. This leaves open a fundamental question: can 1D KPZ universality emerge in a continuous polariton condensate in a planar microcavity, without etched lattices or negative-mass engineering?

In planar microcavities, under non-resonant pumping, realizing such a continuous condensate that exhibits KPZ dynamics is challenging. The condensate is sustained by a reservoir of incoherent excitons, which both feeds and repels the polaritons \cite{carusotto2013quantum}. When the reservoir strongly overlaps the condensate, the associated gain tends to amplify density modulations instead of smoothing them \cite{estrecho2018single}, leading to self-focusing and modulational instabilities rather than a stable extended fluid. At the level of the effective phase dynamics, the reservoir-induced contribution drives the phase diffusion coefficient, $\nu$ negative for realistic parameters, so that phase gradients grow instead of smoothing. This makes it difficult to obtain a long-lived, spatially extended condensate with positive phase diffusion, a prerequisite for KPZ scaling, while preserving a continuous geometry.

\begin{figure}[h!]
    \centering
    \includegraphics[width=1.\linewidth]{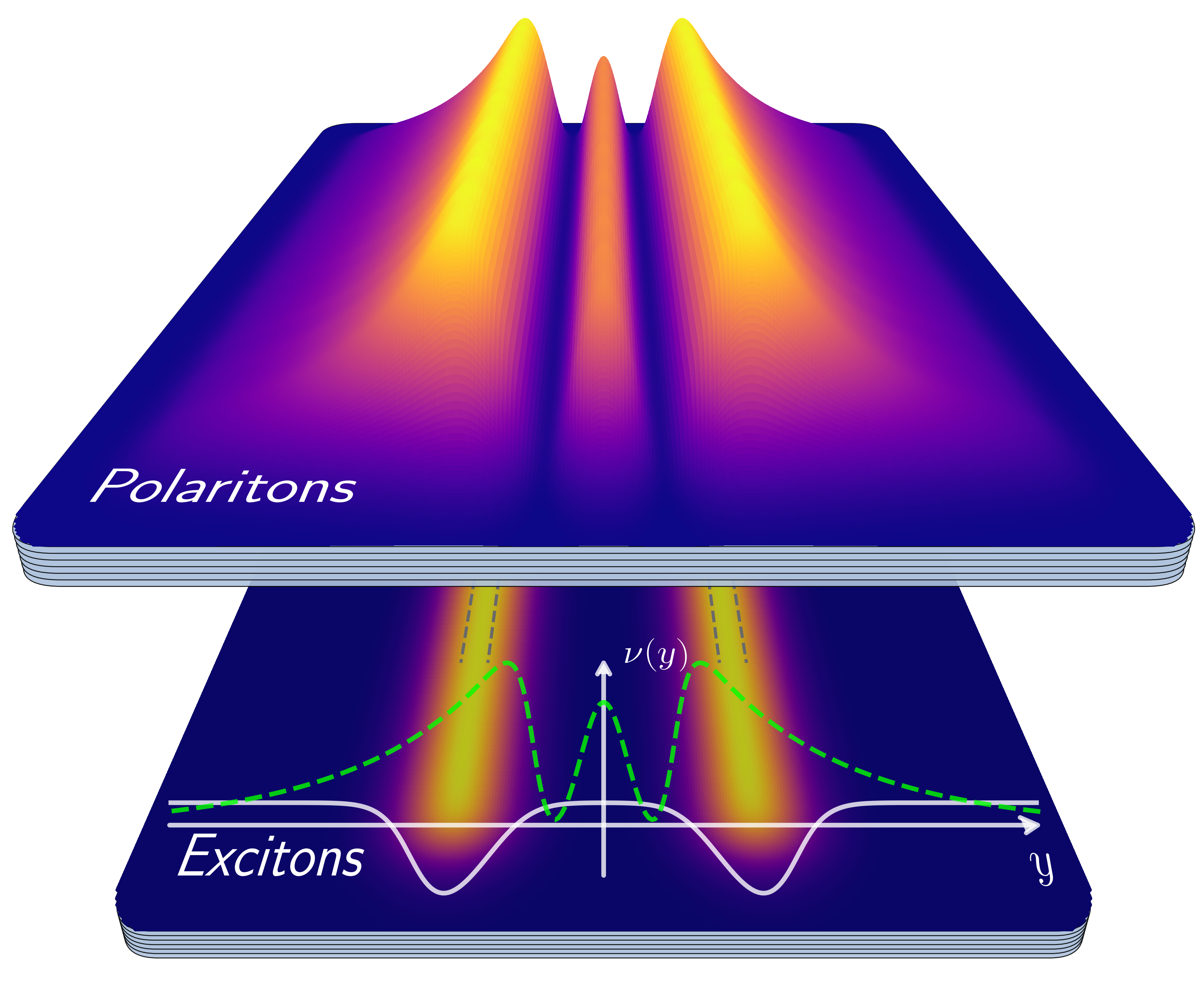}
    \caption{\label{fig:1}
Optically confined polariton condensate formed by two pump stripes. A planar microcavity is nonresonantly pumped in two elongated regions (gray dashed outlines indicate the pump FWHM), creating exciton reservoirs (bottom layer, color map) that feed the lower-polariton branch (LPB). The repulsive reservoir potential confines the condensate to an interference fringe between the stripes with minimal spatial overlap with the reservoirs, yielding a positive effective phase-diffusion coefficient $\nu$ along the unconfined axis $x$ (white curve). The polariton density cross section is indicated by the green dashed line.} 
\end{figure}

In this Letter, we show that KPZ scaling can nevertheless arise in a spatially continuous, quasi-one-dimensional polariton condensate in a planar microcavity, without any etched structures or band-engineering of the dispersion. The key idea is to implement a purely optical, reconfigurable confinement geometry presented in Fig.~\ref{fig:1}: two parallel, elongated non-resonant pumps create a pair of exciton reservoirs (depicted on the bottom colourmap layer) that repel the condensate and generate a standing-wave polariton mode (presented on top colourmap layer) whose intensity maximum is localized in the reservoir-free region between them. This configuration confines the condensate to a quasi-1D stripe that extends over hundreds of microns, while strongly reducing its overlap with the exciton reservoirs. As a result, the central fringe of the condensate experiences a positive and sizable phase diffusion coefficient $\nu$ as shown on the bottom plane of Fig.\ref{fig:1} even for modest high-momentum losses, and remains stable over long times. We show that along the unconfined direction the phase field exhibits all hallmark signatures of the 1D KPZ universality class. The one-point phase fluctuations follow Tracy–Widom statistics of the Gaussian Orthogonal Ensemble \cite{tracy1996orthogonal}, consistent with effectively flat initial conditions. The spatio-temporal phase correlation function obeys the Family–Vicsek scaling form \cite{vicsek1984dynamic} with exponents $\beta \simeq 0.30$ and $\alpha \simeq 0.46$, close to the exact 1D KPZ values, and collapse onto the universal KPZ scaling function \cite{prahofer2004exact}. Together, these results demonstrate that a simple, purely optical confinement scheme in a planar microcavity can realize KPZ universality in a continuous driven–dissipative quantum fluid. The programmable character of the pump profile turns this minimal setup into a reconfigurable analog simulator of non-linear, non-equilibrium universality classes, bridging previous lattice-based realizations with continuous quantum fluids of light.

We utilize the effective mean-field model for the description of polariton condensates, the Gross-Pitaevskii equation \cite{carusotto2013quantum}, modified to account for gain and dissipation. The reservoir-induced stochasticity in the mean field is described by adding the Langevin white noise term, $\xi: ~ \langle\xi^*(\textbf{r}',t')\xi(\textbf{r},t)\rangle=2\xi_0^2\delta(\textbf{r}-\textbf{r}')\delta(t-t')$.

\begin{equation}
    \begin{aligned}
        &\begin{split}
        \scalemath{0.7}{i\hbar\,\partial_t\psi(\mathbf{r},t)=
        \left[
        -\frac{\hbar^2\nabla^2}{2m}
        +\hbar g_{\rm pol}|\psi(\mathbf{r},t)|^2
        +\hbar g\big(n_a(\mathbf{r},t)+n_i(\mathbf{r},t)\big)\right]\psi(\mathbf{r},t)+}\\
        \scalemath{0.7}{+\frac{i\hbar}{2}\left[Rn_a(\mathbf{r},t)-\gamma(\mathbf{k})
        \right]\psi(\mathbf{r},t)
        +i\hbar\,\xi(\mathbf{r},t),}
        \end{split}\\
        & \scalemath{0.85}{\partial_t n_a (\mathbf{r}, t) = -(\gamma_{ex, a} +R|\psi(\mathbf{r}, t)|^2)\,n_a(\mathbf{r}, t)+ W n_i(\mathbf{r}, t),} \\
        & \scalemath{0.85}{\partial_t n_i(\mathbf{r}, t) = -(\gamma_{ex, i} +W)\,n_i(\mathbf{r}, t)+ P(\mathbf{r}),}\\
    \end{aligned}
\end{equation}
where $\psi(\mathbf{r}, t)$ is the polariton wave function, $n_a(\mathbf{r}, t)$ and $n_i(\mathbf{r}, t)$ are the densities of active and inactive exciton resevoirs; $P(\mathbf{r})$ is the incoherent exciton pump. $m$ is the polariton mass, $\gamma(\mathbf{k})$, $\gamma_{ex, a}$ and $\gamma_{ex, i}$ are the polariton loss rate and the exciton loss rates for active and inactive reservoirs, with $\gamma(\mathbf{k})$ $\gg$ $\gamma_{ex, a/i}$. We assume a quadratic dependence of the polariton loss on the wavevector, $\gamma(\mathbf{k}) = \gamma_0 + \tfrac{1}{2}\gamma_2 |\mathbf{k}|^2$, which ensures kinetic-energy dissipation. $R$ is the scattering rate from bottleneck excitons into the polariton condensate, $W$ is the conversion rate from inactive to active excitons. Interactions are described by $g_{pol}$ and $g$, representing polariton--polariton and exciton--polariton interactions, respectively. Exact parameter values used in the simulations are listed in Supplementary Information (SI) I. 

Next we assume a homogeneous in the unconfined direction stationary solution to the above equation and perform the expansion around it \cite{he2015scaling}. Under the assumption that fluctuations of the polariton density are stationary and homogeneous, we arrive at the KPZ equation for the phase of the polariton condensate (see SI ~II for full derivation):

\begin{equation}
    \partial_t \theta(x,t) = \nu\,\partial_x^2 \theta(x,t) + \frac{\lambda}{2}\bigl[\partial_x \theta(x,t)\bigr]^2 + \sqrt{2D}\,\eta(x,t).
    \label{eq:kpz}
\end{equation} 
Here $\eta(x,t)$ is a Gaussian white noise with zero mean and unit variance, $\nu$ is the effective diffusion coefficient, $\lambda$ is nonlinear coupling, and $D$ is the noise amplitude. The phase diffusion and nonlinear coefficients are defined by the microscopic GPE parameters as follows:

\begin{equation}
\begin{aligned}
    \nu = \frac{\hbar}{mR}\left[g_{pol}\frac{\gamma_{ex, a}}{\gamma_0}\frac{P}{\left(1+\frac{\gamma_{ex, i}}{W}\right)P_{\mathrm{th}}} - g \right] + \frac{\gamma_2}{4},\\
    \lambda=-\frac{\hbar}{m} + \frac{\gamma_2}{R}\left[g_{pol}\frac{\gamma_{ex, a}}{\gamma_0} \frac{P}{\left(1+\frac{\gamma_{ex, i}}{W}\right)P_{\mathrm{th}}}-g\right],
\end{aligned}
    \label{eq:nu_expression}
\end{equation}
where $P_{\mathrm{th}} = \gamma_0\gamma_{ex, a}/R$ is the condensation threshold. When the condensate strongly overlaps the reservoir, the repulsive condensate-reservoir interactions $g$ dominates over the polariton–reservoir interaction $g_{pol}$ driving $\nu$ negative. In this regime, density and phase modulations are exponentially amplified, leading to modulational instabilities, filamentation, and rapid vortex proliferation rather than a stable extended fluid \cite{fontaine2022kardar}. In the regions where the polariton fluid is only fed by the in-plane influx of polaritons, however, the reservoir is absent, leading to $g$ effectively equal to zero that leads to positive $\nu$, see Fig.~\ref{fig:1}. The exact details of the proposed configuration together with the stability analysis of the system are provided in SI ~III.

To assess whether the system belongs to the KPZ universality class, we perform the simulations of the full continuous 2D stochastic Gross-Pitaevskii equation and obtain the statistics for the polariton phase field $\theta(x,t)$. We use several diagnostics that probe both one-point and two-point statistics of the phase. A first global observable is the interface roughness, defined as the variance of the phase across the condensate at a given time. For condensate of size $L$ it is expected to grow as $t^{2\beta}$ at early times and to saturate as $L^{2\alpha}$ at late times, with a crossover time scaling as $t_{\mathrm{sat}}\sim L^{z}$. Here $\alpha$ and $\beta$ are the spatial roughness and temporal growth exponents, respectively, and $z$ is the dynamic exponent. For one-dimensional KPZ, their values are: $2\alpha = 1$, $2\beta = 2/3$, and $z = \alpha/\beta = 3/2$ \cite{kardar1986dynamic}. Extensive numerical studies have provided estimates in two dimensions \cite{moser1991numerical, gomes2021kardar}. Although the roughness of the one-point phase-field is not approachable experimentally, here we have extracted the unwrapped phase of the condensate along the unconfined axis and estimated the corresponding critical exponents for sufficiently large lengths, $L$, for KPZ scaling to emerge.  
A detailed analysis of the phase roughness in our system is presented in SI Sec.~IV, wherein for $L\in [175,700]~\mu m$ we obtain $2\alpha_R\approx0.90$ and  $~2\beta_R \approx 0.67$.

\begin{figure}[h!]
    \centering
    \includegraphics[width=0.9\linewidth]{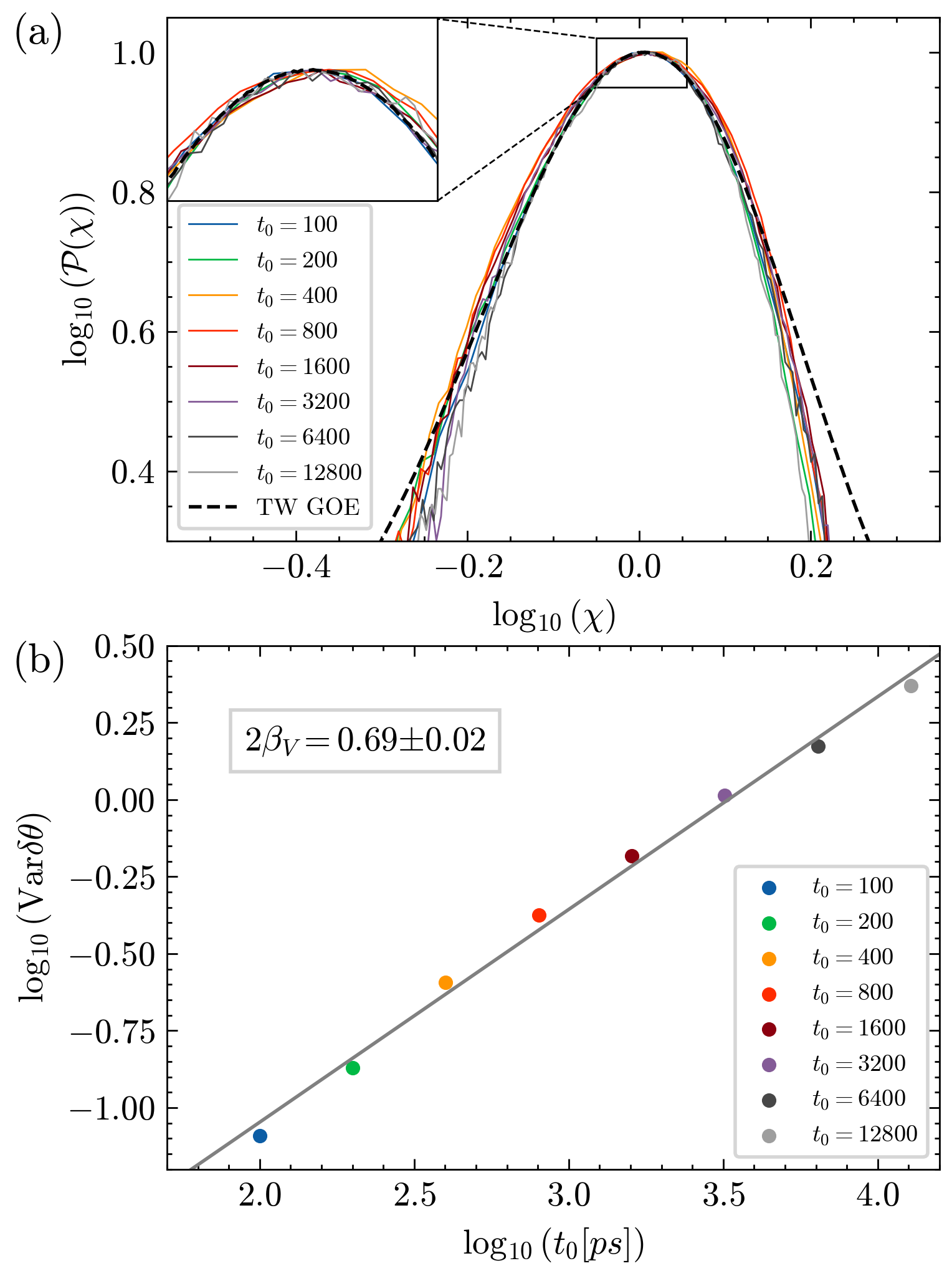}
    \caption{\label{fig:2} 
    Phase-fluctuation statistics in the optically confined condensate.
(a) Probability density of rescaled phase fluctuations
\(\chi=\delta\theta/(t_0)^{1/3}\) evaluated at several initial times \(t_0\), showing collapse onto a single curve. The black dashed line is the Tracy--Widom Gaussian Orthogonal Ensemble distribution (TW GOE), consistent with flat initial conditions.
(b) Variance \(\mathrm{Var}(\delta\theta)\) as a function of \(t_0\); a power-law fit on a double-logarithmic plot yields a growth exponent \(2\beta_V\simeq0.69\) (KPZ: \(2\beta=2/3\)).}
\end{figure}

A more refined hallmark of 1D KPZ universality with flat initial conditions is provided by the asymmetric one-point statistics of the re-scaled phase fluctuations. We consider the fluctuations $\delta \theta(t_0)=\theta(t_0)-\langle \theta(t_0)\rangle$ at a fixed reference time $t_0$  and construct the rescaled variable $\chi=\delta \theta(t_0)/t_0^{1/3}$. KPZ theory predicts that, for flat initial conditions, the probability distribution $\mathcal{P}(\chi)$ converges at long times to the Tracy--Widom distribution associated with the Gaussian Orthogonal Ensemble (TW-GOE) ~\cite{sasamoto2005spatial, ferrari2005determinantal, calabrese2011exact}. In Fig.~\ref{fig:2}a we plot $P(\chi)$ for several times $t_0$ and find that all curves collapse onto a single, time-independent distribution that is well fitted by the TW-GOE form. This is consistent with the effectively flat phase profile of the noiseless steady state, which sets the initial condition for the KPZ dynamics. In Fig.~\ref{fig:2}b we show the variance of $\delta\theta(t_0)$ as a function of $t_0$ in a double-logarithmic scale and extract a growth exponent $2\beta_v \simeq 0.69$, in good agreement with the theoretical KPZ value $2\beta=2/3$.

Beyond one-point statistics, KPZ universality also constrains the full spatio-temporal correlations of the phase field through the Family--Vicsek scaling hypothesis. We characterize these correlations via the two-point phase correlation function

\begin{equation}
\begin{aligned}
  C(\Delta x, \Delta t) = \langle\theta( x_0+\Delta x, t_0+\Delta t)\theta( x_0, t_0)\rangle - \\
  - \langle\theta( x_0+\Delta x, t_0+\Delta t)\rangle\langle \theta( x_0, t_0)\rangle.
\end{aligned}
\end{equation}
where the averages are taken over stochastic realizations and positions $x_0$ along the channel. For a 1D KPZ interface one expects the scaling form

\begin{figure*}[t!]
    \centering
    \includegraphics[width=0.8\linewidth]{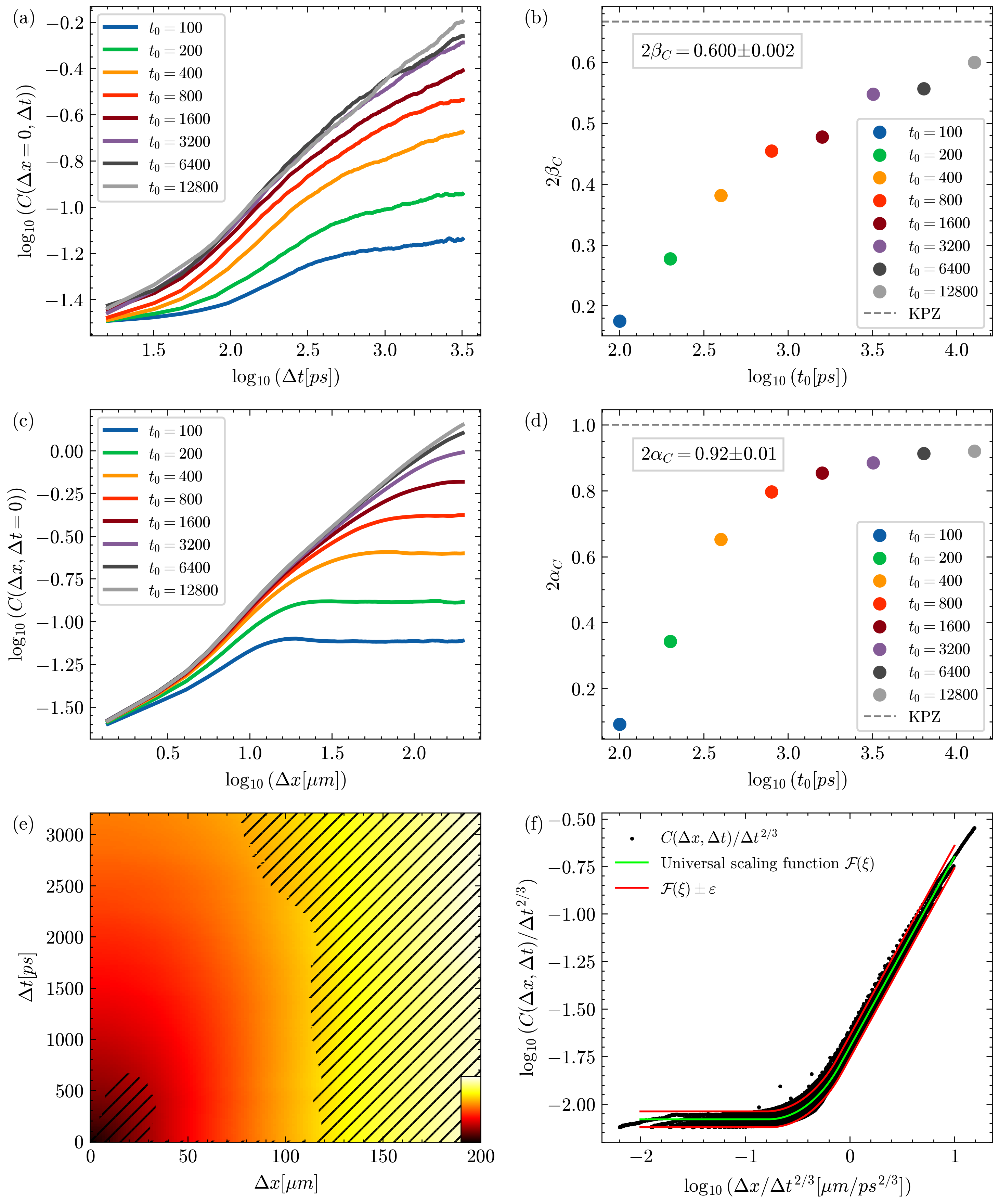}
   \caption{\label{fig:3} 
   Two-point phase correlations and KPZ scaling.
(a) Purely temporal correlations \(C(\Delta x{=}0,\Delta t)\) of the phase, evaluated for different initial times \(t_0\) as a function of time lag \(\Delta t\) (double-logarithmic scale); the slope yields the growth exponent.
(b) Exponent \(2\beta_C\) extracted from (a) versus \(t_0\), approaching the KPZ value \(2\beta=2/3\) (dashed line).
(c) Purely spatial correlations \(C(\Delta x,\Delta t{=}0)\) versus separation \(\Delta x\) for different \(t_0\); the slope yields the roughness exponent.
(d) Exponent \(2\alpha_C\) extracted from (c) versus \(t_0\), approaching the KPZ value \(2\alpha=1\); deviations at late times are limited by finite system size and periodic boundary conditions.
(e) Normalized map of \(C(\Delta x,\Delta t)\); the bright region indicates the scaling window.
(f) Data collapse of the rescaled correlation function onto the universal KPZ scaling form \(\mathcal{F}(\xi)\) with \(\xi=\Delta x/\Delta t^{2/3}\); the green line shows the theoretical \(\mathcal{F}(\xi)\) and the red lines delimit the fitted scaling regime.}
\end{figure*}
\noindent 
\begin{equation}
\label{eq:scaling}
  C(\Delta x, \Delta t) = C_0 (\Delta t)^{2\beta} \
  \mathcal{F}\left(y_0\frac{\Delta x}{(\Delta t)^{1/z}}\right),  
\end{equation}
where $\mathcal{F}$ is a universal scaling function that depends only on the universality subclass, while $C_0$ and $y_0$ are non-universal normalization constants. The asymptotic behaviour of $\mathcal{F}$ encodes the power-law scalings $C(0,\Delta t)\sim \Delta t^{2\beta_c}$ and $C(\Delta x,0)\sim \Delta x^{2\alpha_c}$, so that both the temporal growth exponent $\beta_C$ and the spatial roughness exponent $\alpha_C$ can be extracted from appropriate cuts of $C(\Delta x,\Delta t)$.

In Fig.~\ref{fig:3}a, we show the purely temporal correlations $C(\Delta x=0,\Delta t)$ for several initial times $t_0$, plotted on a double-logarithmic scale. After the system reaches saturation, we observe a clear power-law regime from which we extract a temporal exponent $2\beta_C \simeq 0.60$ at the latest accessible $t_0$, approaching the theoretical KPZ value $2\beta_C=2/3$. The corresponding estimates $2\beta_C(t_0)$ as a function of $t_0$ are summarized in Fig~\ref{fig:3}b, illustrating the gradual convergence towards the universal exponent as the KPZ scaling regime is reached. Figures~\ref{fig:3}c and ~\ref{fig:3}d show the purely spatial correlations $C(\Delta x,\Delta t=0)$ and the extracted roughness exponent $2\alpha_C$, respectively. The build-up of spatial coherence with time leads to an exponent $2\alpha_C \simeq 0.92$ at the latest $t_0$, close to the 1D KPZ value $2\alpha=1$, with the small deviation attributable to finite-size effects, finite cutoff time and the use of periodic boundary conditions. (See SI Sec.~V for a more detailed discussion on contribution of these factors and comparison with other results.)

Finally, the full consistency of our phase dynamics with the KPZ universality class is demonstrated by the collapse of the rescaled two-point correlation function onto the known universal scaling function $\mathcal{F(\xi)}$ of 1D KPZ with flat initial conditions \cite{prahofer2004exact}. In Fig.~\ref{fig:3}e we plot $C(\Delta x,\Delta t)$ as a colour map, where the bright region indicates the domain in $(\Delta x,\Delta t)$ space consistent with the universal scaling. In Fig.~\ref{fig:3}f we rescale $C(\Delta x,\Delta t)$ according to Eq.~\ref{eq:scaling} 
and plot it as a function of the scaling variable $\xi=\Delta x/\Delta t^{2/3}$. The numerical data collapse onto the analytic KPZ scaling function $\mathcal{F(\xi)}$ (blue line) within one standard deviation (red lines), except for two well-understood regions: an early-time crossover near $(\Delta x,\Delta t)=(0,0)$ and a large-$\Delta x$ regime where the finite system size and periodic boundary conditions artificially enhance correlations between opposite points on the ring.

We note that the critical exponents extracted from the roughness behavior ($2\alpha_R\approx0.90, ~2\beta_R \approx 0.67$), the one-point phase variance ($2\beta_V\approx 0.69$) and the two-point correlation function ($2\alpha_C\approx 0.92,~2\beta_C \approx 0.60$) are in good agreement with expected KPZ critical exponents for 1D system. It is also noticeable that the growth exponent obtained from the correlation function is smaller than those extracted through the other methods, due to the finiteness of the cutoff time, available in our simulations. This is also in agreement with previous observations \cite{he2015scaling}. 

In conclusion, we have proposed and numerically analyzed a configuration that realizes Kardar–Parisi–Zhang dynamics in the phase of a spatially continuous polariton condensate of reduced effective dimensionality. Beyond providing a concrete route to observe KPZ scaling, the scheme illustrates the broader potential of optically imprinted polariton condensates as analog simulators of non-equilibrium universality classes. 
In particular, the full reconfigurability of our approach allows future studies of different geometric subclasses of KPZ by shaping the profile of the steady state solution \cite{deligiannis2021accessing}. Alternatively, a similar experiment in curved geometry, when the reservoir is shaped as a ring with a spot in the center, as in \cite{dreismann2014coupled}, can provide crucial insights into the influence of boundary conditions on KPZ physics. Furthermore, the tunability of the effective diffusion coefficient may provide access to a new inviscid Burgers fixed point of the KPZ equation that has recently been predicted by \cite{gosteva2024inviscid, gosteva2025unveiling}. Our approach demonstrates how one of the 2 dimensions of a conventional planar microcavity can be utilized to engineer the stability of the condensate, while keeping the other dimension pure and unconfined for the observation of large scale condensed matter phenomena.

\bibliographystyle{unsrt}
\bibliography{bibliography_main}

\end{document}


\preprint{APS/123-QED}

\title{Supplementary Information: Kardar-Parisi-Zhang physics in optically confined continuous polariton condensates}

\author{Mikhail Misko$^{1,\bullet}$, Natalia Starkova$^{1,\bullet}$, and Pavlos G. Lagoudakis$^{1}$ }
\thanks{Correspondence address: p.lagoudakis@skol.tech\\$^{\bullet}$ M.M. and N.S. contributed equally to this work}

\affiliation{$^1$Hybrid Photonics Laboratory, Skolkovo Institute of Science and Technology, Territory of Innovation Center Skolkovo, Bolshoy Boulevard 30, building 1, 121205 Moscow, Russia}

\date{\today}
                              
\maketitle

\section{Full list of the parameters}

Here, a full list of the parameters that were used in the simulations can be found:

\begin{center}
\begin{tabular}{|c|c|c|}
\hline
Parameter & Units & Value\\
\hline
Planck constant $\hbar$ & meV ps & 0.6591 \\
Effective polariton mass $m$ & meV ps$^2$ $\mu$m$^{-2}$ & 0.28 \\
Polariton decay rate $\gamma_0$ & ps$^{-1}$ & $1/5.5$ \\
Dispersion of polariton decay rate $\gamma_2$ & ps$^{-1}$ $\mu$m$^2$  & $0.1$ \\
Active reservoir decay rate $\gamma_{ex, a}$ & ps$^{-1}$ & $1/{20}$\\
Inactive reservoir decay rate $\gamma_{ex, i}$ & ps$^{-1}$ & $1/{500}$\\
Polariton-polariton interaction strength $g_{pol}$ & ps$^{-1}$ $\mu$m$^2$& 0.005\\
Polariton-reservoir interaction strength $g$/$g_{pol}$ & 1 & 10\\
Rate of scattering $R$ & ps$^{-1}$ $\mu$m$^2$ & 0.1975\\
Rate of conversion $W$ & ps$^{-1}$ & 0.1 \\
Dispersion of Gaussian noise $\xi_0$ & ps$^{-1/2}$ & $0.3$\\
\hline
\end{tabular}
\end{center}

The parameter values used in the simulations are the regular values for a planar GaAs microcavity with embedded InGaAs quantum wells \cite{cilibrizzi2014polariton}. Those values have been used to accurately reconstruct multiple experiments performed on the same sample and are taken from the paper \cite{topfer2020time}.

\section{Mapping to the KPZ equation}

The stochastic generalized GPE, coupled to the equations for active and inactive exciton reservoirs, which we have used to describe our system, looks as follows:

\begin{equation}
    \left\{
    \begin{aligned}
        &i\hbar\,\partial_t\psi(\mathbf{r},t)=
        \left[
        -\frac{\hbar^2\nabla^2}{2m}
        +\hbar g_{\rm pol}|\psi(\mathbf{r},t)|^2
        +\hbar g\big(n_a(\mathbf{r},t)+n_i(\mathbf{r},t)\big)
        +\frac{i\hbar}{2}\Big(Rn_a(\mathbf{r},t)-\gamma(\mathbf{k})\Big)
        \right]\psi(\mathbf{r},t)
        +i\hbar\,\xi(\mathbf{r},t),\\
        &\partial_t n_a (\mathbf{r}, t) = -\left(\gamma_{ex, a} +R|\psi(\mathbf{r}, t)|^2\right)n_a(\mathbf{r}, t)+ W n_i(\mathbf{r}, t),\\
        &\partial_t n_i(\mathbf{r}, t) = -\left(\gamma_{ex, i} +W\right)n_i(\mathbf{r}, t)+ P(\mathbf{r}), \\
    \end{aligned}
    \right.
\end{equation}
where $\gamma(\textbf{k}) = \gamma_0 + \frac12 \gamma_2 |\textbf{k}|^2$, and $\xi$ is Gaussian noise with the following correlation function $\langle \xi(\textbf{r}, t) \xi^{*}(\textbf{r}', t') \rangle = 2 \xi_0^2 \delta(\textbf{r}-\textbf{r}') \delta(t-t')$. 

For the long-time phase dynamics of interest, and on timescales $t\gg \gamma_{{\rm ex},a}^{-1},\gamma_{{\rm ex},i}^{-1}$, the reservoirs relax to a quasi-stationary state. We therefore approximate them by their local steady-state values and set $\partial_t n_{a,i}\approx 0$ to obtain Eq.~(S2). The wavefunction will be written in the following form: $\psi = a e^{i\theta}=ae^{-i\omega t}$. The stable-state solution reads as follows:
\begin{equation}
    \left\{
    \begin{aligned}
        &n_{i,0} = \frac{P}{W+\gamma_{ex, i}}, \\
        &n_{a,0} = \frac{P/\gamma_{ex, a}}{\left( 1+\frac{\gamma_{ex, i}}{W}\right)\left(1+\frac{a_0^2}{n_s}\right)} = n_a(a_0), \\
        &a_0 = \sqrt{n_s p}, \\
        &\omega_0 = g_{pol} a_0^2 + g \left(n_{a, 0}+n_{i, 0}\right),
    \end{aligned}
    \right.
\end{equation}
where we have denoted the pump control parameter $p=P/\left(\left( 1+\frac{\gamma_{ex, i}}{W}\right)P_\mathrm{th}\right)-1$, the threshold pump $P_\mathrm{th} = \gamma_0\gamma_{ex, a}/R$, and the saturation density $n_s=\gamma_{ex, a}/R$. 

We will consider small fluctuations of the wavefunction around the stable state within the rotating frame: $\psi=(a_0+\delta a)e^{i\theta-i\omega_0t}$. We assume that amplitude variations are homogeneous in the long wavelength limit, i.e. $\nabla \delta a=0$, so we arrive at the following two equations for amplitude and phase fluctuations:
\begin{equation}
    \left\{
    \begin{aligned}
        & \delta \dot{a}=-\frac{\hbar}{2m}
        \Delta \theta (a_0+\delta a)-\frac{\gamma_2}{4}
        (\nabla \theta)^2 (a_0+\delta a)-\frac12 R \chi(p)a_0\delta a+ \Re(\xi e^{-i\theta+i\omega_0t}),\\
        &\dot{\theta} = -\frac{\hbar}{2m}(\nabla \theta)^2+\frac{\gamma_2}4 \Delta\theta + (-2g_{pol}a_0+g\chi(p))\delta a+\frac{1}{a_0}\Im(\xi e^{-i\theta+i\omega_0t}),\\
    \end{aligned}
    \right. 
\end{equation}
where we denote $\chi(p)=-\partial n_a/\partial a|_{a_0}=\frac{2P_\mathrm{th}}{\gamma_{ex, a}\sqrt{n_s}}\frac{\sqrt{p}}{1+p}>0$. Since fluctuations in polariton density are subject to restoring force unlike phase fluctuations that proliferate, we decouple the equations assuming $\delta \dot{a}=0$. We arrive at the KPZ equation for the phase of polariton condensate:
\begin{equation}\label{eq:4}
    \dot\theta = \nu\Delta \theta +\frac{\lambda}{2}(\nabla\theta)^2 + \sqrt{2D}\,\eta,
\end{equation}
where $\nu$ is the diffusion coefficient, $\lambda$ is the nonlinearity coefficient, and $\eta$ is the Gaussian noise with the correlation function $\langle \eta(\textbf{r}, t) \eta^{*}(\textbf{r}', t') \rangle = \delta(\textbf{r}-\textbf{r}') \delta(t-t')$. The KPZ coefficients may be derived from the macroscopic GPE parameters as follows:
\begin{equation}
    \left\{
    \begin{aligned}
    &\nu=\frac{\hbar}{m}u + \frac{\gamma_2}4,\\
    &\lambda=-\frac{\hbar}{m} + \gamma_2u, \\
    & D=\frac{\xi_0^2}{2a_0^2}\left(1+4u^2\right),\\
    \end{aligned}
    \right.
\end{equation}
where $u=\frac1R(g_{pol}\frac{\gamma_{ex, a}}{\gamma_0} (1+p)-g)$.

\section{Configuration stability analysis}

\begin{figure}[h!]
    \centering
    \includegraphics[width=0.9\linewidth]{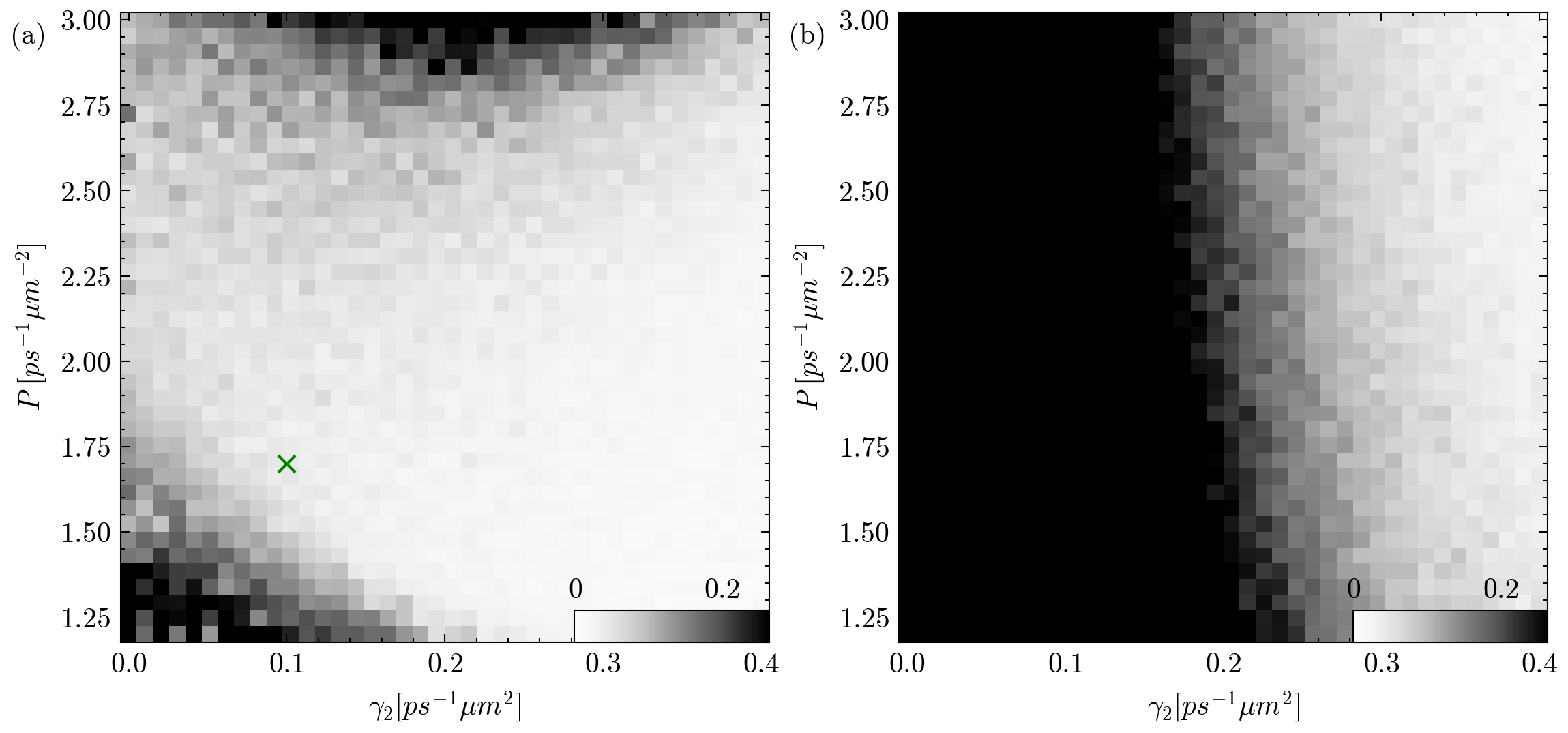}
    \caption{\label{fig:S1} 
Stability analysis of the optically confined condensate. The relative intensity variance along the central stripe vs loss dispersion and pumping power for a) two stripe configuration b) one stripe configuration.  The parameter set used for the KPZ analysis is indicated by the green cross.}   
\end{figure}

We configure the pump in a way to produce two elongated stripes of width $\sim 2.5 \mu m$ and separated $10 \mu m $ apart in order to ensure a bright fringe in the center. To test the validity of our configuration for study of long-time large-scale phase dynamics we vary the $\gamma_2$ coefficient of the polariton losses, as it has been responsible in many previous studies for the stabilisation and diffusion of the condensate. We calculate the relative intensity variance along the central fringe at a large time of $t=10^4$ ps for different values of pumping power and plot it as a colourmap on Fig.~\ref{fig:S1}a. As a control-group experiment, we repeat the same simulation for the case of a single stripe of the same length, and plot the result on Fig.~\ref{fig:S1}b. For the case of the two stripe configuration the confinement stabilizes the condensate, and for a pumping power slightly above threshold there is a window where stability is reached for any value of $\gamma_2$. 

The parameters used for the KPZ analysis are marked by the green cross in Fig.~\ref{fig:S1}. In contrast, for a single stripe a positive $\gamma_2$ is typically required to obtain a sufficiently uniform density profile, which becomes impractical when the condensate substantially overlaps with the reservoir. In the present optically confined configuration, we set $P=2$ and $\gamma_2=0.1$ and verify dynamical stability up to $t=10^{6}\,\mathrm{ps}$.

\section{Roughness of the phase. Saturation time}

\begin{figure}[h!]
    \centering
    \includegraphics[width=0.85\linewidth]{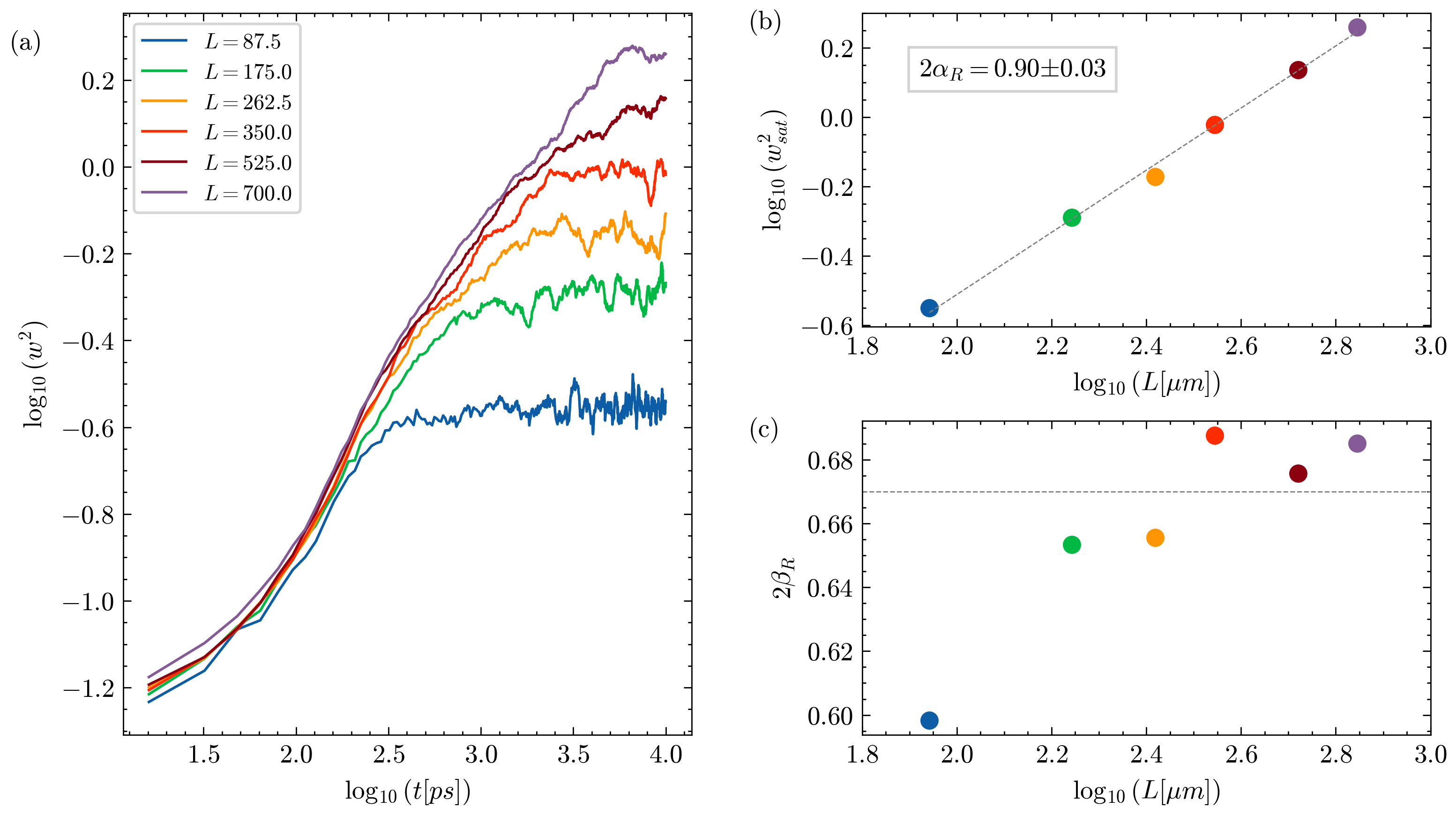}
    \caption{\label{fig:S2} 
    Phase roughness and finite-size scaling.
(a) Roughness \(w^2(t)=\langle [\theta(x,t)-\bar{\theta}(t)]^2\rangle_x\) along the unconfined axis as a function of time for several system sizes \(L\) (double-logarithmic scale).
(b) Saturation roughness \(w_{\mathrm{sat}}\) versus \(L\); a power-law fit yields \(w_{\mathrm{sat}}^2\propto L^{2\alpha_R}\) with \(2\alpha_R\simeq0.9\) (KPZ: \(2\alpha=1\)).
(c) Growth exponent \(2\beta_R\) extracted from the early-time scaling \(w^2(t)\propto t^{2\beta_R}\) as a function of \(L\) (KPZ: \(2\beta=2/3\)).}
\end{figure}

For the selected parameter set and several lengths of the stripes ranging from $L=87.5 \mu m$ to $L=700 \mu m$ with periodic boundaries we plot the extracted roughness of the phase (variance along the spatial coordinate, later averaged over 100 statistical realisations) versus simulation time on Fig.~\ref{fig:S2}a. For each system size we extract the value at which the roughness saturates and plot it 
on Fig.~\ref{fig:S2}b with a dot of respective color in the double-logarithmic scale. We fit the dependence with a line and extract the critical exponent $2\alpha_R\approx 0.9$ which is close to the theoretical value of $1$. We also fit the linearized growth region of roughness and extract the corresponding growth exponent $2\beta_R$ for each system size, and plot it on Fig.~\ref{fig:S2}c.
\clearpage

\section{Discussion on scaling exponents and truncation factors}

Here we analyse the fitting errors that occur when the critical exponents are extracted and compare them to recent experimental and theoretical works in the table below. \\

We notice that the fitting error alone does not describe the divergence of critical exponents from the theoretical values. Since KPZ scaling is a long-time large-size asymptotic of the system, the two main factors that truncate the KPZ exponents are the finite size of the system and the finite simulation times. For example, the values $\beta_C$ and $\alpha_C$, which differ the most from the theoretical values, are obtained by the two-point correlation function analysis after some cutoff time $t_0$, when it is believed that the system is in the saturation regime. From Fig.~3b\&d of the main text it is visible that as $t_0$ increases, the values of the extracted critical exponents approach a plateau that is, however, smaller than the expected theoretical value. This observation signifies that the primary truncating factor in our simulations is the system size, and not the finite cutoff time.

\begin{center}
\begin{tabular}{|c|c|c|}
\hline
Paper & Growth exponent & Roughness exponent\\
\hline
Theoretical values & 
$\begin{aligned}
\beta_{1D}=1/3 \\
\end{aligned}$ &
$\begin{aligned}
\alpha_{1D} = 1/2 \\
\end{aligned}$ \\
(\cite{kardar1986dynamic} 1D, \cite{moser1991numerical, gomes2021kardar} 2D) & $\beta_{2D}\approx 0.241$ & $\alpha_{2D}\approx 0.388$ \\
\hline
\cite{he2015scaling} (1D) & $\beta_{num}=0.317 - 0.307$ & - \\
\cite{deligiannis2021accessing} (1D) & $\beta_{num}=0.30\pm0.01$ & $\alpha_{num}=0.49\pm0.01 $\\
\cite{fontaine2022kardar} (1D) & $\beta_{exp}=0.36\pm0.11$ & $\alpha_{exp}=0.51\pm0.08 $\\
\cite{deligiannis2022kpz} (2D) & $\beta_{num}=0.22\pm0.06$ & $\alpha_{num}=0.36\pm0.04$\\
\cite{widmann2025observation} (2D) & $\beta_{exp}=0.246\pm0.028$ & $\alpha_{exp}=0.417\pm0.034$\\
\hline
Our work & 
$\begin{aligned}
\beta_V = 0.34 \pm 0.01 \\
\beta_C = 0.300 \pm 0.001\\
\beta_R = 0.342 \pm 0.004\\
\end{aligned}$ &
$\begin{aligned}
\alpha_C = 0.46 \pm 0.01 \\
\alpha_R = 0.45 \pm 0.02 \\
\end{aligned}$ \\
\hline
\end{tabular}
\end{center}

\bibliographystyle{unsrt}
\bibliography{bibliography_SI}